\documentclass[journal]{IEEEtran}
\usepackage[english]{babel}
\usepackage[utf8]{inputenc}
\usepackage[T1]{fontenc}
\usepackage{amssymb,array,bm,amsthm}
\usepackage[pdftex]{graphicx}
\usepackage{pbox}
\usepackage{qtree}
\usepackage{ifthen}
\usepackage{tikz}
\usepackage{tabularx}
\usepackage{array}
\usepackage{multirow}
\usepackage{cite}
\usepackage{subcaption}
\usepackage{multirow}
\usepackage{float}
\newtheorem{definition}{Definition}

\newtheorem{theorem}{Theorem}
\newtheorem{property}{Property}

\usepackage[abs]{overpic}

\usepackage[cmex10]{amsmath}

\usepackage{siunitx} % Para alinear por punto decimal
\usepackage{booktabs} % Para mejores líneas en tablas
\usepackage[switch, pagewise]{lineno}
\usepackage[compact]{titlesec}         % you need this package
\titlespacing{\section}{1pt}{1pt}{1pt} % this reduces space between (sub)sections to 0pt, for example
\AtBeginDocument{%                     % this will reduce spaces between parts (above and below) of texts within a (sub)section to 0pt, for example - like between an 'eqnarray' and text
  \setlength\abovedisplayskip{1pt}
  \setlength\belowdisplayskip{1pt}}

\usepackage[hyphens]{url}

\makeatletter
\let\old@ps@headings\ps@headings
\let\old@ps@IEEEtitlepagestyle\ps@IEEEtitlepagestyle
\def\psccfooter#1{%
    \def\ps@headings{%
        \old@ps@headings%
        \def\@oddfoot{\strut\hfill#1\hfill\strut}%
        \def\@evenfoot{\strut\hfill#1\hfill\strut}%
    }%
    \def\ps@IEEEtitlepagestyle{%
        \old@ps@IEEEtitlepagestyle%
        \def\@oddfoot{\strut\hfill#1\hfill\strut}%
        \def\@evenfoot{\strut\hfill#1\hfill\strut}%
    }%
    \ps@headings%
}
\makeatother

\begin{document}
\title{A Cooperative Game-Theoretic Framework for Real-Time Deviation Settlement in Local Energy Communities}
\author{Angel M. González-Rueda, Iago Núñez Lugilde, Jose~L.~Crespo-Vazquez \&  Tarek Alskaif~\IEEEmembership{Senior Member,~IEEE}

\thanks{
Angel M. González-Rueda (Corresponding Author) is with MODESTYA Research Group, Centre for Mathematical Research and Technology of Galicia (CITMAga), Universidade de Santiago de Compostela, Spain (e-mail: angelmanuel.gonzalez.rueda@usc.es). 
Iago Núñez Lugilde is with CITMAga, and SiDOR Research Group, University of Vigo, Spain (e-mail: iago.nunez.lugilde@uvigo.gal).
Jose~L.~Crespo-Vazquez is with University of Vigo, Spain (e-mail: jcrespo@uvigo.es).
Tarek Alskaif is with the Information Technology Group, Wageningen University, The Netherlands (e-mail: tarek.alskaif@wur.nl).}
%\thanks{Send correspondence to Á.~M.~González-Rueda, E-mail: angelmanuel.gonzalez.rueda@usc.es.}% <-this 
\thanks{
This work is part of the R+D+I project grants PID2021-124030NB-C31, and PID2025-170122NB-C33 funded by MCIN/AEI/10.13039/501100011033/ and by ``ERDF A way of making Europe''/EU. This research was also funded by Grupos de Referencia Competitiva ED431C-2021/24 from the Consellería de Cultura, Educación e Universidades, Xunta de Galicia.
}}

% \author{\IEEEauthorblockN{\'Angel M. González-Rueda\IEEEauthorrefmark{1},
% Iago Núñez Lugilde\IEEEauthorrefmark{2},
% Jose~L.~Crespo-Vazquez\IEEEauthorrefmark{3} and 
% Tarek Alskaif\IEEEauthorrefmark{4}}
% \IEEEauthorblockA{
% \IEEEauthorrefmark{1} MODESTYA Research Group and CITMAga, University of Santiago de Compostela, Santiago de Compostela, Spain}
% \IEEEauthorblockA{\IEEEauthorrefmark{2}
% SiDOR Research Group, University of Vigo,  Ourense, Spain}
% \IEEEauthorblockA{\IEEEauthorrefmark{3} 
% University of Vigo, % Vigo, Spain}
%\IEEEauthorblockA{\IEEEauthorrefmark{4}Information Technology Group, Wageningen University, Wageningen, the Netherlands}
% }

\maketitle
\begin{abstract}
This paper proposes a cooperative game-theoretical framework for the financial settlement of real-time deviations in Community-Based Local Energy Markets (CB-LEMs). The real-time settlement process is modeled as a transferable-utility cooperative game, and the resulting Real-Time Local Energy Market (RTLEM) game is shown to be superadditive and balanced. We analyze two game-theoretical allocation mechanisms, the Shapley value and the nucleolus, and evaluate their efficiency, individual rationality, coalitional rationality, and temporal-aggregation properties. We subsequently compare these mechanisms with a settlement rule based on proportional imbalance responsibility. The three mechanisms are efficient and individually rational, but only the nucleolus is guaranteed to satisfy coalitional rationality for every RTLEM instance. The proposed framework is demonstrated using real-world data from a residential energy community in Amsterdam, showing the trade-offs among the three allocation mechanisms in terms of surplus distribution, coalitional stability, interpretability, and computational requirements.

%Initial results indicate that these rules offer advantages over existing proportional allocation methods in terms of stability and equity.

\end{abstract}

%\onecolumn

%\begin{IEEEkeywords}

\vspace{1em}

\section{Introduction}\label{Intro}

Community-Based Local Energy Markets (CB-LEM) represent a promising model for supporting decentralized, renewable-based and user-centric energy systems \cite{sousa2019peer, van2020integrated, faraji2024preference}. In CB-LEMs, community members coordinate their Distributed Energy Resources (DERs) and electricity consumption, enabling the community to optimize its collective energy exchange and interact with the external electricity market at the Point of Common Coupling (PCC) through a representative entity. Indeed, the operational framework of these communities can inherently follow a two-stage market paradigm: a day-ahead (D-1) scheduling phase, where energy exchange commitments are established, and a real-time (RT) balancing phase, where physical injections and withdrawals inevitably deviate from the forecasted schedules due to the stochastic nature of renewable generation and electric demand \cite{visser2024probabilistic}.

Extensive research on LEM has focused on day-ahead (D-1) scheduling and market clearing, including the use of game-theoretic approaches for Peer-to-Peer (P2P) trading \cite{tushar2020}, the construction of consumer coalitions for cost savings \cite{han2018constructing, gramatico2023}, and the design of incentive-compatible market operations \cite{tushar2018peer, mitridati2021design, guo2021online}. Beyond day-ahead scheduling and market clearing, the allocation of costs and economic benefits among community members is essential for ensuring equitable participation and sustaining cooperation. Cooperative game-theoretic approaches have been investigated for allocating distribution network reinforcement costs among photovoltaic installations \cite{park2026game}, distributing revenues from flexibility services among renewable energy community members \cite{faraji2026innovative}, and designing online payoff allocation mechanisms for real-time energy markets \cite{raja2023online}. The work in \cite{alahmed2024dynamic} proposed a dynamic net-metering mechanism for energy communities that guarantees individual and group rationality while aligning members' incentives with community welfare maximization. More related to imbalance settlement, the work in \cite{eriksen2026sharing} investigated the cooperative allocation of imbalance costs and aggregation savings among electricity consumers managed by a common balance responsible party, comparing alternative allocation mechanisms under a two-price imbalance settlement scheme.

However, the ex-post financial settlement of real-time deviations between scheduled and realized energy exchanges within CB-LEMs, particularly the fair and coalitionally stable allocation of the resulting costs and benefits among community members, has received comparatively limited attention \cite{Moret2018, Villar2018}. Individual deviations may partially offset each other, reducing the community's net imbalance settlement costs and generating economic benefits through internal compensation. The fundamental challenge is therefore to distribute these benefits while accounting for members' contributions to collective deviation compensation and ensuring individual and coalitional rationality. Existing approaches include proportional settlement rules, such as the mechanism proposed in \cite{crespo2020community}. While such empirical mechanisms are intuitive and may ensure that individual community members are not worse off than settling independently (i.e., satisfying individual rationality), they treat the community as a homogeneous pool and do not fully capture the marginal contribution of each member to collective benefits. As this paper will demonstrate, such settlements can suffer from a structural vulnerability: they may lead to unstable cross-subsidies. Prosumers whose internal deviations compensate each other may end up subsidizing the imbalance costs of other members with larger deviations. These allocations can violate core stability, creating economic incentives for some sub-coalitions to settle independently rather than remain within the community.

To ensure the long-term socio-economic sustainability of CB-LEMs, the deviation settlement mechanism must be robust against subgroup defection. To address this gap, this paper models the ex-post financial settlement process as a cooperative game, where the value of a coalition reflects its ability to internally balance real-time deviations. We analyze two well-established game-theoretical allocation rules, namely: the Shapley value and the nucleolus, and compare them with the proportional settlement rule proposed in \cite{crespo2020community}, extending the latter to explicitly account for economic terms. By evaluating key axiomatic properties, our analysis reveals how these allocation mechanisms differ in their ability to ensure individual rationality, coalitional stability and equitable benefit sharing.

The main contributions of this work can be summarized as:
\begin{itemize}
    \item We formulate the ex-post financial settlement of real-time deviations in CB-LEMs as a cooperative game, addressing a gap in the literature where this ex-post settlement stage has received limited attention.
    \item We establish key theoretical properties of the resulting cooperative game, ensuring that the settlement process is superadditive and balanced, guaranteeing the existence of coalitionally stable allocations. We further analyze the efficiency, individual rationality and coalitional rationality of the considered settlement rules.
    \item We investigate three settlement mechanisms: the Shapley value, the nucleolus, and the proportional rule proposed in \cite{crespo2020community}. We demonstrate that the proportional rule and the Shapley value can violate core stability and lead to cross-subsidies, while the nucleolus provides a core-stable allocation for the formulated game.
    \item We conduct an illustrative computational study to compare the settlement mechanisms in terms of economic benefit allocation and coalitional stability. We also discuss opportunities to reduce the computational burden of applying the Shapley value and nucleolus.
\end{itemize}

The remainder of this paper is organized as follows. Section~\ref{sec:problem} presents the CB-LEM framework and formulates the real-time deviation settlement problem. Section~\ref{sec3} describes the methodology where the cooperative game-theoretic formulation and the proportional rule are developed and the properties of the allocations are analyzed. Section~\ref{sec:case} presents the numerical evaluation study and discusses the comparative performance of the settlement mechanisms. Finally, Section~\ref{sec:concl} concludes the paper and outlines directions for future research.

\section{Problem description}\label{sec:problem}

In this section, we describe the problem and formally specify the information available for the financial settlement. We adopt a generalized two-stage market paradigm for CB-LEMs, comprising Day-Ahead scheduling and Real-Time balancing, which is a standard approach in recent literature. To contextualize the operational timeline, we take the decentralized model proposed in \cite{crespo2020community} as a representative baseline, which structures this two-stage paradigm into three sequential chronological phases.

In the first phase, the D-1 clearing problem is solved. Regardless of the specific market clearing mechanism, whether it is a stochastic optimization model as in \cite{crespo2020community} or other P2P trading algorithm \cite{van2020integrated}, each community member (typically prosumers) must decide the amount of power to exchange in the LEM. By the end of day D-1, this clearing process determines both the hourly energy price for internal transactions within the CB-LEM on day D, and the binding commitments of each member regarding the quantity of energy to be exchanged in every time slot.

In the second phase, during RT operation on day D, accurate information on actual solar Photovoltaic (PV) generation and demand becomes available. At this point, each community member operates their flexible assets (e.g., energy storage systems) and physical grid exchanges, aiming to adhere as closely as possible to the commitments made in D-1. However, due to the inherent stochasticity of renewable generation and user behavior, deviations inevitably arise between actual physical outcomes and day-ahead schedules.

Consequently, the community faces a collective imbalance at the PCC with the main grid. In the third and final phase, the Community Manager must perform the ex-post financial settlement, ensuring that the costs or revenues derived from this collective deviation are allocated among community members in a fair and efficient manner.

We now formally define the information available to the Community Manager for performing the settlement. Let $\mathcal{N} = \{1, \dots, n\}$ denote the set of agents (i.e., community members), and $\mathcal{T}$ the set of time slots.
\begin{itemize}
	\item From the solution of the Day-Ahead clearing problem, the following information is available:
	\begin{itemize}
		\item $\hat{P}^{\textit{lem}}_{i,t}$: power commitment in the LEM for agent $i$ at time $t$. Since this represents an internal market exchange, the community commitments are perfectly balanced, satisfying $\sum_{i \in \mathcal{N}} \hat{P}^{\textit{lem}}_{i,t} = 0$.
		\item $\lambda^{\textit{lem}}_t$: price in the CB-LEM at time $t$.
	\end{itemize}
	\item From the Real-Time operation, the following information is available:
	\begin{itemize}
        \item $P^{\textit{grid}}_{i,t}$: actual energy exchanged with the grid in real-time.
		\item $\lambda^{\textit{ret,buy}}_t$, $\lambda^{\textit{ret,sell}}_t$: retail buying and selling prices at time $t$. In practice, these prices satisfy $\lambda^{\textit{ret,buy}}_t \ge \lambda^{\textit{lem}}_t \ge \lambda^{\textit{ret,sell}}_t$, ensuring that trading energy within the community is always economically preferable to trading directly with the main grid.
		
	\end{itemize}
\end{itemize}
%It is important to highlight that, when solving the Real-Time problem, each agent aims to maximize the following objective function:
%\begin{equation}
%\sum_{t\in T}\lambda^{\textit{ret,sell}}_t \cdot [\hat{\Delta}^{\textit{lem}}_{i,t}]^+
%- \lambda^{\textit{ret,buy}}_t \cdot [\hat{\Delta}^{\textit{lem}}_{i,t}]^-
%+ \lambda^{\textit{lem}}_t \cdot \hat{P}^{\textit{lem}}_{i,t}
%\label{objfuncRT}
%\end{equation}
It is important to highlight that the individual benefit/cost of agent $i\in \mathcal{N}$ at time $t\in \mathcal{T}$ according to the RT problem is given by:
$$
\begin{aligned}
f^{RT}_t(i) &=
\lambda^{\textit{ret,sell}}_t \cdot [\hat{\Delta}^{\textit{lem}}_{i,t}]^+
- \lambda^{\textit{ret,buy}}_t \cdot [\hat{\Delta}^{\textit{lem}}_{i,t}]^-
+ \lambda^{\textit{lem}}_t \cdot \hat{P}^{\textit{lem}}_{i,t},
\end{aligned}
$$
where\footnote{For any $x\in\mathbb{R}$, $[x]^+=\max\{x,0\}$ and $[x]^-=\max\{-x,0\}$.} $\hat{\Delta}^\textit{lem}_{i,t} = P^\textit{grid}_{i,t} - \hat{P}^\textit{lem}_{i,t}$ denotes the individual deviation of agent $i$ at time $t$ obtained in RT. Note that the first term is the revenue from surplus, the second one the cost due to deficit and the last one the commitment cost.

The Community Manager, having access to all this information, is responsible for performing the final economic settlement. Crucially, while agents are contractually bound to the community for the current operational day, an unfair settlement rule—such as one enforcing cross-subsidies among users—creates strong financial grievances. If members perceive that they could achieve better economic outcomes by operating independently or forming smaller private sub-coalitions, they will have incentives to leave the CB-LEM in the long term. Therefore, the settlement mechanism must not only balance the collective budget but also guarantee coalitional stability to ensure the sustained participation of all members.

\begin{definition}
Let $\mathcal{N} = \{1, \dots, n\}$ be the set of agents, and $\mathcal{T}$ the set of time slots. A settlement rule is a mapping
\[
\psi : \mathcal{N} \times \mathcal{T} \times \mathcal{I} \to \mathbb{R},
\]
where $\psi_{i,t}$ denotes the monetary amount assigned to agent $i \in \mathcal{N}$ at time slot $t \in \mathcal{T}$, given the information $\mathcal{I}$ available to the Community Manager (including day-ahead commitments, real-time deviations, and market prices).

The total settlement for $i$ over all time slots is given as:
\[
\psi_i = \sum_{t \in \mathcal{T}} \psi_{i,t}.
\]
\end{definition}

A natural property for a settlement mechanism is that it must distribute all the benefits or losses among agents, so that the Community Manager does not generate any profit or loss. This ensures that the settlement fully reallocates the economic impact of deviations.

\begin{property}[Efficient]
Let $\psi_{i,t}$ be a settlement rule that assigns to each agent $i$ an amount at time $t$. We say that the settlement rule is \textbf{efficient} at time $t$ if:
\[
\sum_{i\in \mathcal{N}} \psi_{i,t} 
= \lambda_t^{\textit{ret,sell}} \cdot [P_t^{\textit{net}}]^+ 
- \lambda_t^{\textit{ret,buy}} \cdot [P_t^{\textit{net}}]^-,
\]
where $P_t^{\textit{net}} = \sum_{i\in \mathcal{N}} P^{\textit{grid}}_{i,t}$ is the net power deviation of the community at time $t$.%, $[P_t^{\textit{net}}]^+=\max\{P_t^{\textit{net}},0\}$ and $[P_t^{\textit{net}}]^-=\max\{-P_t^{\textit{net}},0\}$.
\end{property}

While efficiency ensures that the total economic impact is fully allocated at the community level, it does not guarantee that individual agents benefit from the distribution proposed. To assess this aspect, we introduce a benchmark corresponding to the case where agents do not cooperate.

\begin{property}[Individual Rationality]
Let $\psi_{i,t}$ be a settlement rule that assigns to each agent $i$ an amount at time $t$. We say that the settlement rule satisfies \textbf{individual rationality} if,  for all $i\in N$,
$$
\psi_{i,t} \;\geq\; f_t^{RT}(i).
$$
\end{property}

This property ensures that no agent receives a lower payoff than in the standalone scenario. 

Individual rationality does not rule out profitable deviations by groups of agents. A coalition may be able to obtain a greater aggregate payoff by settling the deviations of its members separately from the remaining community. This motivates the following coalitional requirement. Given a coalition of agents $S \in 2^\mathcal{N}$, the aggregate deviations these players generate at the community level in a time instant $t\in \mathcal{T}$ are given by:
$
\hat{\Delta}^\textit{lem}_{S,t} = \sum_{i \in S} \hat{\Delta}^\textit{lem}_{i,t}.
$ Based on these deviations, the benefit/cost associated with the coalition $S$ is calculated as:
$$
f^{RT}_{t}(S) = \lambda^\textit{ret,sell}_t \cdot [\hat{\Delta}^\textit{lem}_{S,t}]^{+}
- \lambda^\textit{ret,buy}_t \cdot [\hat{\Delta}^\textit{lem}_{S,t}]^{-}
+ \lambda^\textit{lem}_t \sum_{i \in S} \hat{P}^\textit{lem}_{i,t}.
$$

\begin{property}[Coalitional Rationality]
Let $\psi_{i,t}$ be a settlement rule that assigns to each agent $i$ an amount at time $t$. We say that the settlement rule satisfies  \textbf{coalitional rationality} if,  for all $S\in 2^{\mathcal N}$,
$$
\sum_{i\in S}\psi_{i,t} \;\geq\; f_t^{RT}(S).
$$
\end{property}

Coalitional rationality implies individual rationality by applying the condition to every singleton coalition $S=\{i\}$.

\section{Methodology}\label{sec3}

This section presents the cooperative game-theoretical methodology used to address the ex-post settlement of real-time deviations. We first formulate the Real-Time Local Energy Market (RTLEM) game and study its  properties. We then introduce the Shapley value and the nucleolus as two point solutions of the cooperative game. These mechanisms are compared with a settlement rule based on proportional imbalance responsibility.

A \textit{cooperative game} is a pair $(\mathcal{N}, v)$, where $\mathcal{N}$, the set of players, is a finite and nonempty subset of the natural numbers, and $v : 2^{\mathcal{N}} \rightarrow \mathbb{R}$ is the \textit{characteristic function}, satisfying $v(\emptyset) = 0$. A \textit{coalition} is any element  $S \in 2^\mathcal{N}$. In general, $v(S)$ represents the maximum value that coalition $S$ can guarantee independently of the players in $\mathcal{N} \setminus S$. For any $x \in \mathbb{R}^{\mathcal{N}}$ and $S\in2^{\mathcal N}$, let $
x(S) = \sum_{i \in S} x_i
$. We denote by $G^{\mathcal{N}}$ the set of all cooperative games with $\mathcal{N}$ as the set of players, and write $v \in G^{\mathcal{N}}$ instead of $(\mathcal{N}, v)$. For simplicity, we write $v(i_1, \ldots, i_s)$ instead of $v(\{i_1, \ldots, i_s\})$.

We say that an allocation $x \in \mathbb{R}^\mathcal{N}$ is \textit{efficient} for a game $v \in G^\mathcal{N}$ if $x(\mathcal{N}) = v(\mathcal{N})$. An allocation $x \in \mathbb{R}^\mathcal{N}$ is \textit{individually rational} if $x_i \geq v(i)$ for all $i \in \mathcal{N}$, meaning that each player receives at least as much as they could secure on their own. The set of all individually rational and efficient allocations is called the \textit{imputation set} and is defined as $I(v) = \bigl\{ x \in \mathbb{R}^\mathcal{N} : x(\mathcal{N}) = v(\mathcal{N}),\ x_i \geq v(i)\ \text{for all } i \in \mathcal{N} \bigr\}$. An allocation $x \in \mathbb{R}^\mathcal{N}$ is \textit{stable} if $x(S) \geq v(S)$ for all $S \in 2^\mathcal{N}$. This condition ensures that no coalition can unilaterally improve its outcome without the need for cooperation from other coalitions. The set of all stable allocations in the game $v$ is $C(v) = \bigl\{ x \in I(v) : x(S) \geq v(S)\ \text{for all } S \in 2^\mathcal{N} \bigr\}$ and is called the \textit{core} of $v$. A game $v \in G^\mathcal{N}$ is \textit{balanced} if its core is nonempty.

A game $v \in G^\mathcal{N}$ is called \emph{additive} if $v(S) = \sum_{i \in S} v(i)$ for all $S \in 2^\mathcal{N}$. %In this case, the game $v$ is fully characterized by the vector $a = \left( v(i) \right)_{i \in N} \in \mathbb{R}^N$. 
%For simplicity, we use the same symbol to denote both the additive game and its corresponding vector.
Let $S, R \in 2^\mathcal{N}$. A game $v \in G^\mathcal{N}$ is said to be %\emph{monotonic} if $v(S) \leq v(R)$ whenever $S \subseteq R$,
\emph{superadditive} if $v(S) + v(R) \leq v(S \cup R)$ whenever $S \cap R = \emptyset$.%, and \emph{convex} \cite{shapley1971} if $v(S \cup \{i\}) - v(S) \leq v(R \cup \{i\}) - v(R)$ for all $S \subset R \subset \mathcal{N} \setminus \{i\}$. 

%Convexity implies superadditivity. Nevertheless, superadditivity does not imply monotonicity. Also, an important subclass of games with a nonempty core is the class of convex games.

A \textit{point solution} on $G^\mathcal{N}$ is a function $\phi: G^\mathcal{N} \rightarrow \mathbb{R}^\mathcal{N}$ that assigns to each game $v \in G^\mathcal{N}$ a unique allocation vector of $\mathbb{R}^\mathcal{N}$.  Let $\phi$ be a point solution, if $\phi(v) \in C(v)$ for every $v \in G^\mathcal{N}$, then $\phi$ is said to be a \textit{stable} solution. Given two games $v,w\in G^\mathcal{N}$, if $\phi(v)+\phi(w)=\phi(v+w)$, then $\phi$ is said to be \textit{additive}.

\subsection{Modeling the Cooperative Game}

In this section, we model the problem defined in Section \ref{sec:problem} as a cooperative game. We study its properties and various definitions for allocating the cost or benefit generated by being part of the CB-LEM, where the set of players corresponds to the prosumers of the LEM, $\mathcal{N} = \{1, 2, \dots, n\}
$. 
%For each agent $i\in\mathcal{N}$, we have the following information for each time instant $t\in \mathcal{T}$:
%\begin{itemize}
%    \item $\hat{P}^\textit{lem}_{i,t}$: energy commitment acquired in the CB-LEM on day D-1.
%    \item $\lambda^\textit{lem}_{t}$: energy price in the CB-LEM set on day D-1.
%    \item $\hat{\Delta}^\textit{lem}_{i,t} = P^\textit{grid}_{i,t} - \hat{P}^\textit{lem}_{i,t}$: individual deviation of agent $i$ at time $t$ obtained in Real-Time (RT).
%\end{itemize}

%Given an agent $i\in \mathcal{N}$ and a time $t\in \mathcal{T}$, the individual benefit/cost associated according to the RT problem is given by:
%$$
%\begin{aligned}
%f^{RT}_t(i) &=
%\lambda^{\textit{ret,sell}}_t \cdot [\hat{\Delta}^{\textit{lem}}_{i,t}]^+
%- \lambda^{\textit{ret,buy}}_t \cdot [\hat{\Delta}^{\textit{lem}}_{i,t}]^-
%+ \lambda^{\textit{lem}}_t \cdot \hat{P}^{\textit{lem}}_{i,t},
%\end{aligned}
%$$
%where the first term is the revenue from surplus, the second one the cost due to deficit and the last one the commitment cost.

To approach this as a cooperative game, we need to define a characteristic function for any time instant $t\in\mathcal T$, $v_t: 2^\mathcal{N} \rightarrow \mathbb{R}$, which assigns to each coalition $S \in 2^\mathcal{N}$ a value $v_t(S)$. For a given coalition $S \in 2^\mathcal{N}$, we define $v_t(S)$ as the payoff that the members of coalition $S$ can obtain by jointly settling their aggregate real-time deviation, conditional on their previously determined day-ahead commitments and the market prices.
Then, we can define the cooperative game associated with this situation, called \textit{Real-Time Local Energy Market} (RTLEM) game.

\begin{definition}
Given a time instant $t\in \mathcal{T}$, the RTLEM game is a cooperative game defined by $v_t\in G^{\mathcal{N}}$, where $\mathcal{N} = \{1, 2, \dots, n\}$ is the set of prosumers (players) participating in the market and $v_t: 2^\mathcal{N} \rightarrow \mathbb{R}$ is the characteristic function that assigns to each coalition $S \in 2^\mathcal{N} $ the aggregate payoff that the members of $S$ can obtain by jointly settling their realized deviations at time $t$, conditional on the day-ahead commitments and market prices:
    $$
    v_t(S) = f_t^{RT}(S).
    $$
\end{definition}

 Note that, for each $t\in\mathcal{T}$, $$v_t(\mathcal{N})=\lambda_t^{\textit{ret,sell}} \cdot [P_t^{\textit{net}}]^+-\lambda_t^{\textit{ret,buy}} \cdot [P_t^{\textit{net}}]^-.$$

%\noindent where $P_t^{\textit{net}}$ is the net deviation of the community at time $t\in \mathcal{T}$, defined by $P_t^{\textit{net}} = \sum_{i \in \mathcal{N}} P^{\textit{grid}}_{i,t} = \sum_{i \in \mathcal{N}} (P^{\textit{grid}}_{i,t}-\hat{P}^{lem}_{i,t}) =\sum_{i \in \mathcal{N}} \hat{\Delta}^{lem}_{i,t}$. 
Therefore, if the community's deviation is positive, $v_t(\mathcal{ })$ corresponds to the profit obtained from selling the surplus to the retailer; and if it is negative, it corresponds to the cost associated with purchasing energy from the grid.

Now, we show some properties of the RTLEM game. First, we show that the game is superadditive. In a superadditive game, the union of two coalitions yields a higher total value than their separate actions, reinforcing the incentive to cooperate. 

%However, the game is not monotonic, as we will show in the example of Section \ref{sec4}. When agents exhibit heterogeneous imbalance positions (e.g., some agents have deficits while others have surpluses), the aggregation of deviations may reduce the total benefit of certain sub-coalitions. As a result, it is possible that for some $S \subset R$, $v(R) < v(S)$. However, this lack of monotonicity does not contradict the economic rationale for cooperation. What ultimately drives participation is not the monotonicity of the characteristic function, but whether individual agents are better off within the grand coalition compared to operating independently. In this context, the relevant condition is individual rationality, which ensures that each agent receives at least its standalone benefit $v(\{i\})$. Even in non-monotonic settings, it is possible to construct allocations such that all agents benefit from cooperation, as we will see in Subsection \ref{subsec:rules}.

\begin{theorem}\label{th:superadditive}
The RTLEM game is superadditive for any time instant $t\in \mathcal{T}$.
\begin{proof}
    Let $S,R\in 2^\mathcal{N}$ such that $S\cap R=\emptyset$ and $t\in{\mathcal T}$. We prove that $v_t(S\cup R)\geq v_t(S)+v_t(R)$. On the one hand,

\begin{align*}
    &v_t(S\cup R)\\&=\lambda^\textit{ret,sell}_t \cdot [\hat{\Delta}^\textit{lem}_{S\cup R,t}]^{+}
- \lambda^\textit{ret,buy}_t \cdot [\hat{\Delta}^\textit{lem}_{S\cup R,t}]^{-}
+ \lambda^\textit{lem}_t \sum_{i \in S\cup R} \hat{P}^\textit{lem}_{i,t},
\end{align*}

and, on the other hand,

\begin{align*}
    &v_t(S)+v_t(R)\\&=\lambda^\textit{ret,sell}_t \cdot [\hat{\Delta}^\textit{lem}_{S,t}]^{+}
- \lambda^\textit{ret,buy}_t \cdot [\hat{\Delta}^\textit{lem}_{S,t}]^{-}
+ \lambda^\textit{lem}_t \sum_{i \in S} \hat{P}^\textit{lem}_{i,t}\\&+\lambda^\textit{ret,sell}_t \cdot [\hat{\Delta}^\textit{lem}_{R,t}]^{+}
- \lambda^\textit{ret,buy}_t \cdot [\hat{\Delta}^\textit{lem}_{R,t}]^{-}
+ \lambda^\textit{lem}_t \sum_{i \in R} \hat{P}^\textit{lem}_{i,t}.
\end{align*}

Therefore,

\begin{align*}
    &v_t(S\cup R)- v_t(S)-v_t(R)\\&=\lambda^\textit{ret,sell}_t \cdot \big([\hat{\Delta}^\textit{lem}_{S\cup R,t}]^{+}-[\hat{\Delta}^\textit{lem}_{S,t}]^{+}-[\hat{\Delta}^\textit{lem}_{R,t}]^{+}\big)\\&-\lambda^\textit{ret,buy}_t \cdot \big([\hat{\Delta}^\textit{lem}_{S\cup R,t}]^{-}-[\hat{\Delta}^\textit{lem}_{S,t}]^{-}-[\hat{\Delta}^\textit{lem}_{R,t}]^{-}\big).
\end{align*}

We have to show that $v_t(S\cup R)- v_t(S)-v_t(R)\geq 0$. If $\hat{\Delta}^\textit{lem}_{S,t}\geq 0$ and $\hat{\Delta}^\textit{lem}_{R,t}\geq 0$, $\hat{\Delta}^\textit{lem}_{S\cup R,t}\geq 0$ and the result is straightforward. So, assume that $\hat{\Delta}^\textit{lem}_{S,t}\geq 0$, $\hat{\Delta}^\textit{lem}_{R,t}\leq 0$, and $\hat{\Delta}^\textit{lem}_{S\cup R,t}\geq 0$. In such a case,

\begin{align*}
    &v_t(S\cup R)- v_t(S)-v_t(R)\\&=\lambda^\textit{ret,sell}_t \cdot \big(\hat{\Delta}^\textit{lem}_{S\cup R,t}-\hat{\Delta}^\textit{lem}_{S,t}\big)-\lambda^\textit{ret,buy}_t \cdot \big(\hat{\Delta}^\textit{lem}_{R,t}\big)\\&=\hat{\Delta}^\textit{lem}_{R,t}\cdot\big(\lambda^\textit{ret,sell}_t-\lambda^\textit{ret,buy}_t\big)\geq 0,
\end{align*}

\noindent since $\lambda^\textit{ret,buy}_t\geq \lambda^\textit{ret,sell}_t$. The remaining cases are analogous.

\end{proof}
\end{theorem}

Next, we show that the game is balanced. In a balanced game, there always exists at least one stable allocation that no coalition has an incentive to block; in other words, the core is nonempty. 

\begin{theorem}
The RTLEM game is a balanced game for any time instant $t\in \mathcal{T}$.

 \begin{proof} Fix a time instant $t\in\mathcal{T}$. According to the Bondareva--Shapley theorem \cite{bondareva1963some}, to prove balancedness, it is sufficient to prove that \begin{equation} \sum_{\emptyset\neq S\subseteq {\mathcal N}} \mu_S v_t(S) \leq v_t({\mathcal N}) \label{eq:Bondareva-Shapley-condition} \end{equation} for every balanced collection $\left(\mu_S\right)_{\emptyset\neq S\subseteq {\mathcal N}}$ of nonnegative weights. Let $\left( \mu_S \right)_{\emptyset\neq S\subseteq N}$ be an arbitrary balanced collection such that $\mu_S\geq 0$ for every $\emptyset\neq S\subseteq {\mathcal N}$ and $\sum_{\substack{S\subseteq {\mathcal N}\\ i\in S}} \mu_S = 1$ for every $i\in {\mathcal N}.$ Define the real-time deviation settlement function $g_t:\mathbb{R}\rightarrow\mathbb{R}$ as $g_t(z) = \lambda_{t}^{\textit{ret,sell}}[z]^+ - \lambda_{t}^{\textit{ret,buy}}[z]^-. $
 
 Using this notation, the characteristic function can be written as $$v_t(S) = g_t\left( \hat{\Delta}_{S,t}^{\textit{lem}} \right) + \lambda_{t}^{\textit{lem}} \sum_{i\in S} \hat{P}_{i,t}^{\textit{lem}}. $$

 We now evaluate the left-hand side of the Bondareva--Shapley condition, \begin{align} \sum_{\emptyset\neq S\subseteq {\mathcal N}} \mu_S v_t(S) &= \sum_{\emptyset\neq S\subseteq {\mathcal N}} \mu_S g_t\left( \hat{\Delta}_{S,t}^{\textit{lem}} \right) \nonumber\\ &\quad+ \lambda_{t}^{\textit{lem}} \sum_{\emptyset\neq S\subseteq {\mathcal N}} \mu_S \sum_{i\in S} \hat{P}_{i,t}^{\textit{lem}}. \label{eq:balanced-sum-decomposition} \end{align}
 
 We first consider the term associated with the day-ahead commitments. We have \begin{align*} \sum_{\emptyset\neq S\subseteq {\mathcal N}} \mu_S \sum_{i\in S} \hat{P}_{i,t}^{\textit{lem}} &= \sum_{\emptyset\neq S\subseteq {\mathcal N}} \mu_S \sum_{i\in {\mathcal N}} \mathbf{1}_{\{i\in S\}} \hat{P}_{i,t}^{\textit{lem}} \nonumber\\ &= \sum_{i\in {\mathcal N}} \hat{P}_{i,t}^{\textit{lem}} \sum_{\emptyset\neq S\subseteq {\mathcal N}} \mu_S \mathbf{1}_{\{i\in S\}} \nonumber\\ &= \sum_{i\in {\mathcal N}} \hat{P}_{i,t}^{\textit{lem}} \sum_{\substack{S\subseteq {\mathcal N}\\i\in S}} \mu_S= \sum_{i\in {\mathcal N}} \hat{P}_{i,t}^{\textit{lem}}.  \end{align*} 
 
 Moreover, as the day-ahead LEM commitments are balanced,  \begin{equation} \lambda_{t}^{\textit{lem}} \sum_{\emptyset\neq S\subseteq {\mathcal N}} \mu_S \sum_{i\in S} \hat{P}_{i,t}^{\textit{lem}} = 0. \label{eq:balanced-commitments-zero} \end{equation}

   We next consider the term associated with real-time deviations of \eqref{eq:balanced-sum-decomposition}. It is easy to show that, for any $\alpha\geq 0$ and $z\in\mathbb{R}$, $g_t(\alpha z)=\alpha g_t(z)$. Also, by the same sign-based argument used in Theorem \ref{th:superadditive}, for any $x,y\in\mathbb{R}$, $g_t(x+y)\geq g_t(x)+g_t(y).$ Therefore,

\begin{align} &\sum_{\emptyset\neq S\subseteq {\mathcal N}} \mu_S g_t\left( \hat{\Delta}_{S,t}^{\textit{lem}} \right) = \sum_{\emptyset\neq S\subseteq {\mathcal N}} g_t\left( \mu_S \hat{\Delta}_{S,t}^{\textit{lem}} \right) \nonumber\\&\leq g_t\left( \sum_{\emptyset\neq S\subseteq {\mathcal N}} \mu_S \hat{\Delta}_{S,t}^{\textit{lem}} \right)\nonumber= g_t\left( \sum_{\emptyset\neq S\subseteq {\mathcal N}} \mu_S \sum_{i\in S} \hat{\Delta}_{i,t}^{\textit{lem}} \right)\nonumber\\&= g_t\left( \sum_{i\in {\mathcal N}} \hat{\Delta}_{i,t}^{\textit{lem}} \sum_{\substack{S\subseteq {\mathcal N}\\i\in S}} \mu_S\right)= g_t(\hat{\Delta}_{{\mathcal N},t}^{\textit{lem}}).\label{eq:balanced-deviation-bound} \end{align} 

Substituting \eqref{eq:balanced-commitments-zero} and \eqref{eq:balanced-deviation-bound} into \eqref{eq:balanced-sum-decomposition} gives \begin{align} \sum_{\emptyset\neq S\subseteq N} \mu_S v_t(S) &\leq g_t\left( \hat{\Delta}_{N,t}^{\textit{lem}} \right). \label{eq:balanced-condition-before-efficiency} \end{align} 

Using the definition of $g_t$ and the balanced day-ahead commitments, the value of the grand coalition is \begin{align} v_t(N) &= g_t\left( \hat{\Delta}_{N,t}^{\textit{lem}} \right) + \lambda_{t}^{\textit{lem}} \sum_{i\in N} \hat{P}_{i,t}^{\textit{lem}} = g_t\left( \hat{\Delta}_{N,t}^{\textit{lem}} \right). \label{eq:grand-coalition-value-balanced} \end{align} Thus, from \eqref{eq:balanced-condition-before-efficiency} and \eqref{eq:grand-coalition-value-balanced}, \begin{equation*} \sum_{\emptyset\neq S\subseteq N} \mu_S v_t(S) \leq v_t(N),\end{equation*} and we can conclude that $v_t$ is balanced.

\end{proof}
\end{theorem}

Now we will prove that if all individual deviations in time $t\in \mathcal{T}$ have the same sign, then the RTLEM game is additive. In that case, the core is a singleton, $C(v_t)=\{(v_t(i))_{i\in \mathcal{N}}\}$.

\begin{theorem}\label{theorem:aditive}
    Given a time instant $t\in \mathcal{T}$ in which, for all $i\in \mathcal{N}$, $\hat{\Delta}^\textit{lem}_{i,t}\geq 0$ or $\hat{\Delta}^\textit{lem}_{i,t}\leq 0$, then the RTLEM game is additive.
\end{theorem}
\begin{proof}
Let $S\in2^\mathcal{N}$ and $t\in \mathcal{T}$, we aim to show that $v_t(S)=\sum_{i\in S}v_t(i)$ if $\hat{\Delta}^\textit{lem}_{i,t}\geq 0$ for all $i\in N$ (the opposite case is analogous). In such a case, $[\hat{\Delta}^{\textit{lem}}_{i,t}]^+=\hat{\Delta}^{\textit{lem}}_{i,t}$ and $[\hat{\Delta}^{\textit{lem}}_{i,t}]^-=0$ for all $i\in N$ and then, $
\hat{\Delta}^\textit{lem}_{S,t} = \sum_{i \in S} \hat{\Delta}^\textit{lem}_{i,t}=\sum_{i \in S} [\hat{\Delta}^\textit{lem}_{i,t}]^+
$. Therefore, $[\hat{\Delta}^\textit{lem}_{S,t}]^+=\hat{\Delta}^\textit{lem}_{S,t}$, $[\hat{\Delta}^\textit{lem}_{S,t}]^-=0$, and

\begin{align*}
    v_t(S)&=f^{RT}_{t}(S)\\& = \lambda^\textit{ret,sell}_t \cdot [\hat{\Delta}^\textit{lem}_{S,t}]^{+}
- \lambda^\textit{ret,buy}_t \cdot [\hat{\Delta}^\textit{lem}_{S,t}]^{-}
+ \lambda^\textit{lem}_t \sum_{i \in S} \hat{P}^\textit{lem}_{i,t}\\& = \lambda^\textit{ret,sell}_t \cdot \sum_{i \in S} \hat{\Delta}^\textit{lem}_{i,t}
+ \lambda^\textit{lem}_t \sum_{i \in S} \hat{P}^\textit{lem}_{i,t}\\& = \sum_{i\in S}f^{RT}_{t}(i)=\sum_{i\in S} v_t(i).
\end{align*}
\end{proof}

Finally, we define a game that models the end-of-day allocation by taking into account all the deviations that occurred throughout the day at each time instant $t\in \mathcal{T}$.

\begin{definition}
    The daily LEM game is a cooperative game defined by $v\in G^{\mathcal{N}}$, where $\mathcal{N}$ is the set of prosumers participating in the market and $v$ is the characteristic function defined by $$v=\sum_{t\in{\mathcal T}}v_t,$$
    \noindent being $v_t\in G^{\mathcal{N}}$ the RTLEM game for time instant $t\in\mathcal{T}$.
\end{definition}

Since finite sums preserve superadditivity and balancedness, the daily LEM game is also superadditive and balanced. Furthermore, it is an additive game if all individual deviations in any time $t\in\mathcal{T}$ have the same sign, that is, if for all $t\in\mathcal{T}$, $v_t$ is additive.

\subsection{Settlement Allocation Methods}\label{subsec:rules}

Now, we introduce three settlement rules. The Shapley value and the nucleolus are point solutions defined on the cooperative game, whereas the proportional rule is a data-driven allocation mechanism based on individual deviations and proportional responsibility factors.

\subsubsection{Shapley value}

We begin by defining the Shapley value. To do so, we introduce the concept of \textit{marginal contributions}. Let $v\in G^{\mathcal{N}}$ be a cooperative game and $\pi: \mathcal{N} \to \{1, \dots, |\mathcal{N}|\}$ be a permutation of the players in $\mathcal{N}$. The vector of marginal contributions associated with the order $\pi$ is defined as
$
m^\pi(v) = (m^\pi_i(v))_{i \in \mathcal{N}} \in \mathbb{R}^\mathcal{N},
$
where
$
m^\pi_i(v) = v(\text{Pre}^\pi(i) \cup \{i\}) - v(\text{Pre}^\pi(i)),
$
and $\text{Pre}^\pi(i) = \{ j \in \mathcal{N} : \pi(j) < \pi(i) \}$ is the set of players preceding $i$ in the order $\pi$.

\begin{definition}

The \emph{Shapley value} \cite{shapley} $Sh(v)$, is defined as the average of the marginal contribution vectors over all possible permutations of the players. Thus, the Shapley value of the game $v \in G^\mathcal{N}$ is given by:
$$
Sh(v) = \frac{1}{|\Pi^N|} \sum_{\pi \in \Pi^N} m^\pi(v),
$$
where $\Pi^N$ is the set of all permutations of $N$.
\end{definition}

The Shapley value can be interpreted as follows: suppose players arrive in a random order, and each one receives their marginal contribution upon arrival. Then, the Shapley value is the expected marginal-contribution vector under this random arrival process. Note that, the Shapley value is a settlement rule which is efficient since, for each $t\in\mathcal{T}$, $$\sum_{i\in\mathcal{N}}Sh_i(v_t)=v_t(\mathcal{N})=\lambda_t^{\textit{ret,sell}} \cdot [P_t^{\textit{net}}]^+-\lambda_t^{\textit{ret,buy}} \cdot [P_t^{\textit{net}}]^-.$$

Since the RTLEM game is superadditive, every marginal contribution of player $i$ is at least $v_t(i)$. Consequently, the Shapley value is individually rational and belongs to the imputation set. Therefore, as a settlement rule, it satisfies individual rationality, but, as we will see in Section \ref{sec:case}, it may fail coalitional rationality. Moreover, for any additive game $v\in G^\mathcal{N}$, $Sh_i(v)=v(i)$ for any $i\in \mathcal{N}$.

\subsubsection{The nucleolus}
Now, we continue with the definition of the nucleolus. The key concept behind this rule is the \emph{excess} of a coalition. Let $v\in G^{\mathcal{N}}$ be a cooperative game, and let $x \in \mathbb{R}^\mathcal{N}$ be an allocation, for any coalition $S \in 2^\mathcal{N}$, the excess of $S$ at $x$ is defined by $e(S, x) = v(S) - x(S).$
%where $x(S) = \sum_{i \in S} x_i$ is the total allocation received by the members of $S$. 
If $x(S) < v(S)$, the coalition $S$ is dissatisfied, as its members could potentially achieve a better outcome by acting independently. Thus, the excess $e(S, x)$ quantifies the ``complaint'' of coalition $S$ against the allocation $x$. The larger the excess, the more dissatisfied the coalition. Note that, for every efficient allocation $x$, $e(\mathcal{N}, x) = e(\emptyset, x) = 0$. For a given allocation $x$, we define the vector of excesses by $\theta(x) = \left( e(S_1, x), e(S_2, x), \dots, e(S_{2^{|\mathcal{N}|}}, x) \right),
$ where the coalitions are ordered so that
$
\theta_1(x) \geq \theta_2(x) \geq \cdots \geq \theta_{2^{|\mathcal{N}|}}(x).
$

\begin{definition}

The \emph{ nucleolus} \cite{ nucleolus} of the game $v\in G^\mathcal{N}$ is the set of allocations in the imputation set $I(v)$ that lexicographically minimize the vector of excesses:
$$
\text{Nu}(v) = \left\{ x \in I(v) : \theta(x) \leq_L \theta(y) \quad \text{for all } y \in I(v) \right\},
$$
where $\leq_L$ denotes the lexicographic order.
\end{definition}

If $I(v)\neq \emptyset$, the nucleolus exists and is unique. Similarly to the Shapley value, the nucleolus is efficient by construction. Moreover, because every RTLEM game is balanced, its nucleolus belongs to the core. Thus, no coalition can obtain a strictly greater aggregate payoff by leaving the grand coalition. Therefore, it satisfies coalitional rationality and, therefore, individual rationality.  Also, for any additive game $v\in G^N$, $Nu_i(v)=v(i)$ for any $i\in N$. 

\subsubsection{Proportional rule}

In addition to the two cooperative game-theoretical solutions, we consider the proportional rule as a practical benchmark. Unlike the Shapley value and the nucleolus, this mechanism is not defined as a point solution on the space of cooperative games. The rule is inspired by \cite{crespo2020community}. The idea is to allocate each prosumer’s participation in the CB-LEM according to their individual deviation and the aggregate community deviation. The procedure consists of the following steps.

First, we compute the share of each agent’s deviation with respect to the community total, distinguishing between positive and negative deviations. Formally, we define the coefficients:
\begin{equation*}
\rho^\textit{up}_{i,t} = \frac{[\hat{\Delta}^\textit{lem}_{i,t}]^+}{\sum_{j \in \mathcal{N}} [\hat{\Delta}^\textit{lem}_{j,t}]^+}, 
\qquad
\rho^\textit{down}_{i,t} = \frac{[\hat{\Delta}^\textit{lem}_{i,t}]^-}{\sum_{j \in \mathcal{N}}[\hat{\Delta}^\textit{lem}_{j,t}]^-}.
\end{equation*}

Second, we calculate the net aggregate exchange of the community at each time step $t \in \mathcal{T}$, $P^{\textit{net}}_t = \sum_{i \in \mathcal{N}} P^{\textit{grid}}_{i,t}$. This net power deviation represents the community-level deviation to be settled. Each agent’s participation in the CB-LEM is then adjusted proportionally to their share of this deviation, so that energy exchanged in the opposite direction to the community’s overall deviation is fully assigned as LEM trading. This ensures that participation in the LEM always benefits the agents: at worst, each must compensate for their own deviation, while deviations from others may reduce the amount to be covered, thereby increasing the value of cooperation. The adjusted participation in the LEM for agent $i$ at time $t$ is thus:
\begin{equation}
P^{\textit{lem}}_{i,t} = P^{\textit{grid}}_{i,t} - \rho^\textit{up}_{i,t} \cdot [P^{\textit{net}}_t]^+ + \rho^\textit{down}_{i,t} \cdot [P^{\textit{net}}_t]^-.
\label{eq:lem}
\end{equation}

Finally, this proportional energy distribution is translated into monetary terms, giving the following settlement rule.
\begin{definition}
The \textit{proportional rule} for a time instant $t\in \mathcal{T}$ assigns to each $i \in \mathcal{N}$:
\begin{equation*}
\pi_{i,t} =  \lambda^{\textit{lem}}_t \cdot P^{\textit{lem}}_{i,t} 
+  \lambda^{\textit{ret,sell}}_{t} \cdot [\Delta^{\textit{lem}}_{i,t}]^+ 
- \lambda^{\textit{ret,buy}}_{t} \cdot [\Delta^{\textit{lem}}_{i,t}]^-,
\end{equation*}
\noindent where the assigned deviation is given by 
$\Delta_{i,t}^{\textit{lem}} = P^{\textit{grid}}_{i,t}-P^{\textit{lem}}_{i,t}$.
\end{definition}

We now establish some key properties of the proposed rule. First, we show that this rule is efficient.
\begin{theorem}
The proportional rule $\pi_{i,t}$ is efficient.
\end{theorem}

\begin{proof}
Expanding the total cost and applying the market clearing condition $\sum_{i \in \mathcal{N}} P^{\textit{lem}}_{i,t}=0$ yields:
\[
    \sum_{i \in \mathcal{N}} \pi_{i,t} = \lambda_t^{\textit{ret,sell}} \sum_{i \in \mathcal{N}} \left[ \Delta_{i,t}^{\textit{lem}} \right]^+ - \lambda_t^{\textit{ret,buy}} \sum_{i \in \mathcal{N}} \left[ \Delta_{i,t}^{\textit{lem}} \right]^-.
\]
By definition of the allocation rules, individual deviations $\Delta_{i,t}^{\textit{lem}}$ share the same sign as the net community deviation $P_t^{\textit{net}}$. Because the responsibility coefficients $\rho_{i,t}^{\textit{up}}$ and $\rho_{i,t}^{\textit{down}}$ sum to 1 over $\mathcal{N}$, the aggregate deviation components simplify directly to $\sum_{i \in \mathcal{N}} [\Delta_{i,t}^{\textit{lem}}]^\pm = [P_t^{\textit{net}}]^\pm$. Substituting these identities into the expanded cost yields the claimed result.
\end{proof}

The following result demonstrates that the proportional rule ensures that no agent obtains a lower payoff than they would by operating in isolation. 

\begin{theorem}
The proportional rule satisfies individual rationality. %, i.e.,
%\begin{equation}
%    \pi_{i,t}^{sett} \ge f_t^{RT}(i), \quad \forall i \in \mathcal{N}.
%\end{equation}
\end{theorem}

\begin{proof}
Assume without loss of generality that the net community deviation is positive ($P_t^{\textit{net}} > 0$). If $P_t^{\textit{net}} = 0$, no aggregate deviation is assigned to any agent, and individual rationality follows directly from the assumed price ordering. We analyze the two possible cases for the deviation of an agent $i\in\mathcal N$:

\noindent \textbf{Case 1: The agent deviates in the opposite direction ($\hat{\Delta}_{i,t}^{lem} \le 0$).}

In this case, $[P^{\textit{net}}_t]^{+} = P^{\textit{net}}_t, [P^{\textit{net}}_t]^{-} = 0,$ so the LEM participation of agent $i$ reduces to
\[
P^{\textit{lem}}_{i,t} = P^{\textit{grid}}_{i,t} - \rho^\textit{up}_{i,t}\cdot P^{\textit{net}}_t.
\]

Since $\hat{\Delta}^\textit{lem}_{i,t} = P^\textit{grid}_{i,t} - \hat{P}^\textit{lem}_{i,t} < 0$, 
we have $[\hat{\Delta}^\textit{lem}_{i,t}]^+ = 0$ and $[\hat{\Delta}^\textit{lem}_{i,t}]^- = \hat{P}^\textit{lem}_{i,t} - P^\textit{grid}_{i,t}$.  
It follows that $\rho^\textit{up}_{i,t} = 0$, so the LEM participation simplifies to 
\[
P^{\textit{lem}}_{i,t} = P^{\textit{grid}}_{i,t}.
\]

According to the proportional rule, the assigned payoff is therefore
\[
\pi_{i,t} = \lambda^{\textit{lem}}_t \cdot P^{\textit{grid}}_{i,t}.
\]

On the other hand, the RT objective function of the agent is given by
\[
f^{RT}_t(i) 
= - \lambda^{\textit{ret,buy}}_t \cdot (\hat{P}^\textit{lem}_{i,t}-P^\textit{grid}_{i,t}) 
+ \lambda^{\textit{lem}}_t \cdot \hat{P}^{\textit{lem}}_{i,t}.
\]

We want to show that $\pi_{i,t} \geq f^{RT}_t(i)$, so
\begin{eqnarray*}
& &\lambda^{\textit{lem}}_t \cdot P^{\textit{grid}}_{i,t} \geq
- \lambda^{\textit{ret,buy}}_t \cdot (\hat{P}^\textit{lem}_{i,t}-P^\textit{grid}_{i,t}) 
+ \lambda^{\textit{lem}}_t \cdot \hat{P}^{\textit{lem}}_{i,t}, \\[0.5em]
& \Leftrightarrow & \quad
\lambda^{\textit{ret,buy}}_t \cdot (\hat{P}^\textit{lem}_{i,t}-P^\textit{grid}_{i,t}) 
\geq \lambda^{\textit{lem}}_t \cdot (\hat{P}^{\textit{lem}}_{i,t} - P^{\textit{grid}}_{i,t}), \\[0.5em]
& \Leftrightarrow & \quad
(\hat{P}^\textit{lem}_{i,t}-P^\textit{grid}_{i,t}) \cdot 
(\lambda^{\textit{ret,buy}}_t-\lambda^{\textit{lem}}_t) \geq 0.
\end{eqnarray*}

Since by assumption $\hat{P}^\textit{lem}_{i,t} \geq P^\textit{grid}_{i,t}$ and retail buying prices satisfy $\lambda^{\textit{ret,buy}}_t \geq \lambda^{\textit{lem}}_t$, the inequality holds. 

The case $P^{\textit{net}}_t \leq 0$ with $\hat{\Delta}^\textit{lem}_{i,t} \geq 0$ is analogous.

\noindent \textbf{Case 2: The agent deviates in the same direction ($\hat{\Delta}_{i,t}^{\textit{lem}} > 0$).} The proof for this scenario follows a similar algebraic deduction and is omitted for brevity.

\end{proof}

Although the proportional rule protects isolated individuals, the following theorem shows that it can violate the coalitional rationality property.

\begin{theorem}
The proportional rule may fail to satisfy coalitional rationality. %There exist operational scenarios where a coalition of prosumers $S \subset \mathcal{N}$ receives a joint payoff strictly lower than what they would obtain by settling their deviations independently from the rest of the community:
%\begin{equation}
 %   \sum_{i \in S} \pi_{i,t}^{\textit{sett}} < f_t^{RT}(S)
%\end{equation}
\end{theorem}

\begin{proof} 
This result is illustrated in the numerical study in Section~\ref{sec:case}.
\end{proof}

%This fundamental flaw—the enforcement of cross-subsidies that penalize balanced subgroups to cover extreme individual deviations—creates a strong incentive for community fragmentation. To ensure long-term sustainability, the settlement mechanism must be robust not only at the individual level but for any possible sub-coalition. This naturally motivates the use of cooperative game theory, where concepts such as Core Stability are explicitly designed to prevent subgroup defection.

\subsection{Computational aspects}

While the cooperative game-theoretical framework proposed in this paper is general, the exact computation of the allocation methods may become challenging as the number of community members increases. Indeed, the exact computation of the Shapley value and the nucleolus generally requires processing an exponential number of coalitions, which may become prohibitive for large communities. By contrast, the proportional settlement rule can be computed directly from the individual deviations and the aggregate community imbalance, without enumerating all possible coalitions. 

Nevertheless, the structure of the RTLEM game simplifies the computation of the three allocation methods in some cases. According to Theorem~\ref{theorem:aditive}, if all individual deviations have the same sign at a time instant $t\in\mathcal{T}$, the RTLEM game is additive. Moreover, it follows that the final LEM exchange assigned by the proportional settlement rule coincides with the day-ahead commitment: $ P_{i,t}^{\textit{lem}} = \widehat{P}_{i,t}^{\textit{lem}}.$  As a result, the proportional rule assigns each agent its singleton worth $v_t(i), i\in\mathcal{N}.$ Hence, when all individual deviations have the same sign, the three allocation methods produce the same allocation:  $v_t(i)$,  for all $i\in\mathcal{N}.$ Their computation is therefore immediate.%, and nontrivial cooperative-game calculations are required only for time slots containing individual deviations of both signs. 

The allocation methods also differ in their temporal-aggregation properties. The daily LEM game is defined as the sum of the time-specific RTLEM games. Since the Shapley value is additive across games, its daily allocation satisfies $$ Sh(v) = Sh\left( \sum_{t\in\mathcal{T}}v_t \right) = \sum_{t\in\mathcal{T}}Sh(v_t). $$

Therefore, the same daily Shapley allocation can be obtained either by computing the Shapley value of the aggregated daily game or by summing the time-specific Shapley allocations. By contrast, the nucleolus is not additive across games. Consequently, computing the nucleolus of the daily LEM game and summing the nucleolus of the time-specific RTLEM games represent two different settlement procedures. 

The proportional rule has substantially lower computational requirements. Unlike the Shapley value and the nucleolus, the computation of the proportional rule does not require enumerating the $2^n-1$ nonempty coalitions. Its computational cost grows linearly with the number of agents for each time instant. Since the proportional rule is defined from time-specific market data, its daily allocation could be obtained by summing the settlements computed separately for every time instant. Nevertheless, this temporal aggregation should not be interpreted as additivity in the cooperative game-theoretical sense. %The direct economic settlement is a time-specific allocation mechanism rather than a solution mapping defined on the space of transferable-utility games. 

\section{Numerical Evaluation and Discussions}\label{sec:case}

To illustrate the proposed framework, we consider a real-world case study based on a residential neighborhood in Amsterdam, The Netherlands. The simulation relies on data from ten households, all of which act as prosumers, over the course of a single day. The dataset is adapted from \cite{crespo2020community}, which provides detailed information about each prosumer, including their day-ahead commitments, real-time deviations, and the final allocation of energy exchanges in the LEM, accounting for both planned commitments and actual real-time trades. To compute the Shapley value and the nucleolus, we implemented these calculations using \texttt{TUGLab} \cite{TUGlab}. %, an open-source R package developed for the analysis and implementation of cooperative game-theoretic tools. Since the community consists of ten households, the computation of the rules is straightforward, and no computational issues arise.

We organize the numerical analysis into two settings. The first contains homogeneous deviations, with all prosumers deviating in the same direction. The second contains heterogeneous deviations, with both positive and negative individual deviations.

\subsection{Homogeneous Deviations}

We first consider a time instant at which all prosumers exhibit deviations of the same sign. In this case, the RTLEM game is additive (see Theorem~\ref{theorem:aditive}), and the three settlement rules studied in this paper coincide. Thus, each prosumer $i\in\mathcal{N}$ receives the payoff

$$
f^{RT}_t(i)=
\lambda^{\textit{ret,sell}}_t \cdot [\hat{\Delta}^{\textit{lem}}_{i,t}]^+
- \lambda^{\textit{ret,buy}}_t \cdot [\hat{\Delta}^{\textit{lem}}_{i,t}]^-
+ \lambda^{\textit{lem}}_t \cdot \hat{P}^{\textit{lem}}_{i,t}.
$$

As an illustrative example, consider $t=2$, for which  Table~\ref{tab:lem_agents_hom} reports the relevant data. The corresponding prices are $\lambda^{\textit{ret,sell}}_2=5.46$, $\lambda^{\textit{ret,buy}}_2=12.62$, and $\lambda^{\textit{lem}}_2=12.46$.

\begin{table}[h] \centering \footnotesize \sisetup{ table-align-text-post = false,   round-mode = places, round-precision = 2 } \begin{tabular}{c S[table-format=-2.2] S[table-format=-2.2] S[table-format=-2.2] S[table-format=-2.2]} \toprule Prosumer & {$\hat{P}^{\textit{lem}}_{i,2}$} & {$P^{\textit{grid}}_{i,2}$} & {$\hat{\Delta}^\textit{lem}_{i,2}$} & {$f_2^{RT}(i)$} \\ \midrule 1 & 0.67 & 0.65 & -0.02 &8.10 \\ 2 & -1.18 & -1.59 & -0.41&-19.88 \\ 3 & 0.00 & -0.04 & -0.04&-0.50 \\ 4 & -0.24 & -0.24 & 0.00& -2.99\\ 5 & 0.79 & 0.41 & -0.38& 5.05 \\ 6 & 1.07 & 0.60 & -0.47&7.40 \\ 7 & -0.51 & -0.52 & -0.01&-6.48 \\ 8 & -0.33 & -0.39 & -0.06&-4.87 \\ 9 & 0.00 & -0.04 & -0.04&-0.50 \\ 10 & -0.27 & -0.31 & -0.04&-3.87 \\ \bottomrule \end{tabular} \caption{Data for the 10 prosumers participating in the LEM at time instant $t=2$.} \label{tab:lem_agents_hom} \end{table}

Since all deviations are negative, $[\hat{\Delta}^{\textit{lem}}_{i,2}]^+=0$ and $[\hat{\Delta}^{\textit{lem}}_{i,2}]^-=-\hat{\Delta}^{\textit{lem}}_{i,2}$ for every $i\in\mathcal{N}$. Therefore, the three settlement rules considered in this paper coincide and assign to each prosumer $i\in\mathcal{N}$ the payoff

$$
f_2^{RT}(i)=12.62\cdot \hat{\Delta}^{\textit{lem}}_{i,2}+ 12.46 \cdot \hat{P}^{\textit{lem}}_{i,2}.
$$

\subsection{Heterogeneous Deviations}

We next analyze a time instant characterized by both positive and negative individual deviations.  We consider, for example, the time instant $t=20$, for which the data reported in Table~\ref{tab:lem_agents_het1} are obtained. The corresponding prices are $\lambda^{\textit{ret,sell}}_{20}=4.45$, $\lambda^{\textit{ret,buy}}_{20}=10.95$, and $\lambda^{\textit{lem}}_{20}=10.95$. %Note that, $\sum_{i\in{\mathcal{N}}}\hat{\Delta}^\textit{lem}_{i,20}=1.39>0$. In this case, the community exhibits a positive aggregate deviation, which is settled by selling the resulting net surplus to the retailer.

\begin{table}[h] \centering \footnotesize \sisetup{ table-align-text-post = false,   round-mode = places, round-precision = 2 } \begin{tabular}{c S[table-format=-2.2] S[table-format=-2.2] S[table-format=-2.2] S[table-format=-2.2]} \toprule Prosumer & {$\hat{P}^{\textit{lem}}_{i,20}$} & {$P^{\textit{grid}}_{i,20}$} & {$\hat{\Delta}^\textit{lem}_{i,20}$} & {$f_{20}^{RT}(i)$} \\ 
\midrule 
1 & -1.69 & -2.27 & -0.58 &-24.86 \\
2 & 13.09 & 11.25 & -1.84&123.19 \\ 
3 & 0.00 & 0.06 & 0.06&0.27 \\
4 & -1.43 & 0.17 & 1.60& -8.54\\
5 & -3.04 & -3.81 & -0.77& -41.72 \\ 
6 & 0.00 & 0.00 & 0.00&0.00 \\ 
7 & -3.18 & -0.85 & 2.33&-24.45 \\
8 & -2.19 & -1.76 & 0.53&-22.72 \\ 
9 & 0.00 & -0.01 & -0.01&-0.11 \\
10 & -1.46 & -1.39 & 0.07&-15.68 \\ \bottomrule \end{tabular} \caption{Data for the 10 prosumers participating in the LEM at time instant $t=20$.} \label{tab:lem_agents_het1} \end{table}

Since deviations have mixed signs, the game is non-additive and the three settlement rules can yield different allocations. As shown in Table~\ref{tab:lem_agents_het1}, the prosumers' deviations can be either positive or negative. The amount to be distributed is

$$
v(\mathcal{N})=\lambda^{\textit{ret,sell}}_{20}\cdot\sum_{i\in {\mathcal N}}\hat{\Delta}^\textit{lem}_{i,20}=6.14.
$$

Therefore, at this time, the community has enough energy to meet its own needs and has a surplus that can be sold to the grid. The resulting proceeds are then distributed among the prosumers. Table~\ref{table:t20} shows the allocations obtained under the three settlement rules.\footnote{Minor discrepancies between the displayed column totals and $v_{20}({\mathcal N})$ are due to rounding. All allocation rules are efficient when evaluated using unrounded values.}

\begin{table}[h] \centering \footnotesize \sisetup{ table-align-text-post = false,   round-mode = places, round-precision = 2 } \begin{tabular}{c S[table-format=-2.2]  S[table-format=-2.2] S[table-format=-2.2]} \toprule Prosumer & {$Sh_i(v_{20})$} & {$Nu_i(v_{20})$} & {$\pi_{i,20}$} \\ 
\midrule 
1 & -22.62 & -21.08 & -24.86\\
2 & 130.46 & 135.16 & 123.19 \\ 
3 & 0.39 & 0.28 & 0.56\\
4 & -5.88 & -8.54 & -1.29\\
5 & -38.80& -36.73 & -41.72\\ 
6 & 0.02 & 0.01 & 0.02 \\ 
7 & -20.02 & -24.45& -13.87 \\
8 & -21.77 & -22.72 & -20.33 \\ 
9 & -0.09 & -0.06& -0.11 \\
10 & -15.54 & -15.67 & -15.36 \\ \bottomrule \end{tabular} \caption{Allocations proposed by the three settlement rules at time instant $t=20$.} \label{table:t20} \end{table}

%\begin{figure}[h]
%    \centering
%    \includegraphics[width=1\linewidth]{Distribution_t20.eps}
%    \caption{Distribution of the cost for the Shapley value (blue), the nucleolus (red) and the economic settlement (green) for the RTLEM game in $t=20$.}
 %   \label{fig:t20}
%\end{figure}

\begin{figure}[h]
    \centering
    \includegraphics[width=1\linewidth]{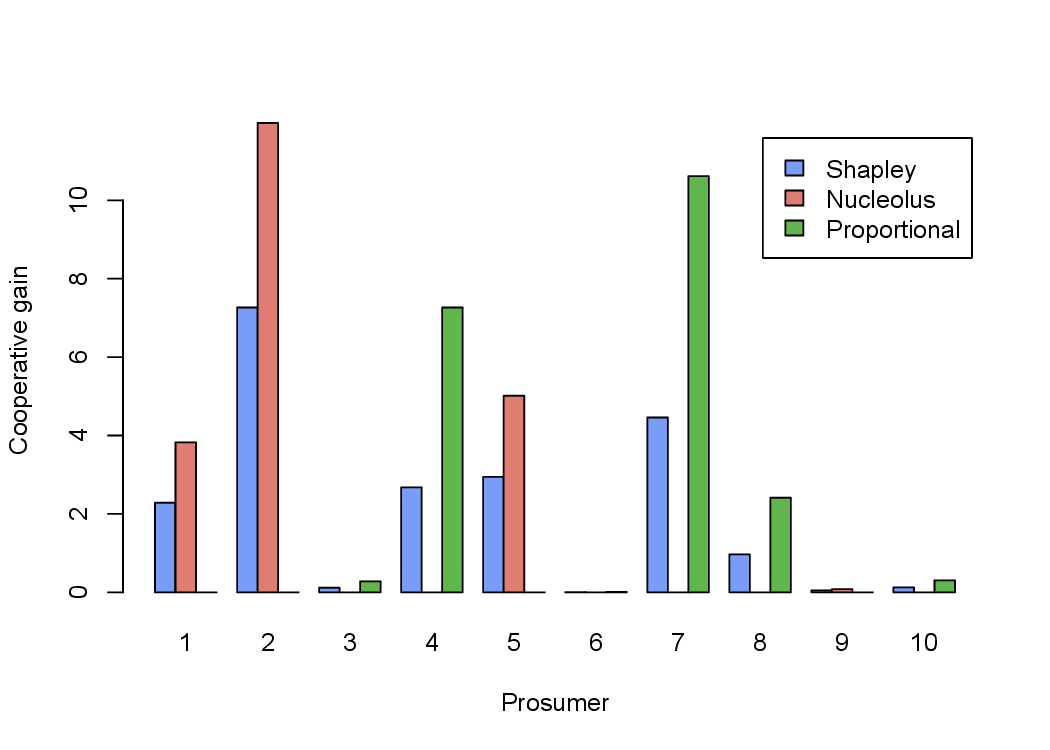}
    \caption{Cooperative gain $g_{i,t}^r=x_{i,t}^r-v_t(i)$ assigned to each prosumer by the settlement rules at $t=20$.}
    \label{fig:Dif_t20}
\end{figure}

Although the three rules distribute the same aggregate payoff, they differ in how the cooperative surplus is allocated among prosumers. For allocation rule $r$, the cooperative gain assigned to prosumer $i\in\mathcal{N}$ at time $t\in\mathcal{T}$ is defined as $g_{i,t}^r=x_{i,t}^r-v_t(i)$, where $x_{i,t}^r$ denotes the allocation under rule $r$, and $v_t(i)$ is the payoff that prosumer $i$ could obtain by settling independently. 
 Figure~\ref{fig:Dif_t20} reports the cooperative gain received by each prosumer at $t=20$. A value equal to zero indicates that the payoff from the rule matches exactly $v_{20}(i)$; positive values show that the prosumer receives more than their stand-alone payoff; and no negative cooperative gains appear, indicating that all three allocation rules satisfy individual rationality in this instance. The Shapley value assigns a strictly positive cooperative gain to every prosumer in this instance. By contrast, the nucleolus and the proportional rule leave several prosumers exactly at their stand-alone payoff and concentrate a larger share of the cooperative surplus among a smaller group of participants.

%The Shapley value yields a strictly positive cooperative gain for every agent, whereas the nucleolus and the direct economic settlement leave several agents exactly at their stand-alone worth. In this way, the figure highlights the differences between the solutions more clearly: the nucleolus and the direct economic settlement concentrate the cooperative surplus among a smaller number of agents, leaving several participants at or close to their individual worth, $v_{20}(i)$. In contrast, the Shapley value provides positive deviations above 
%the individual worth to a greater number of prosumers, although these deviations are generally smaller. This broader distribution of benefits may serve as a stronger incentive for widespread continued participation in the community. 

To complement the analysis, we examine the distribution of cooperative gains throughout the day.\footnote{Cooperative gains were rounded to five decimal places before being classified as strictly positive or zero, and we only consider time instants with heterogeneous deviations.} The Shapley value assigns a strictly positive gain in 99.58\% of the prosumer-time observations, compared with 70.83\% for the proportional rule and 30.00\% for the nucleolus. The corresponding medians are 0.17, 0.025, and 0, respectively. Therefore, the Shapley value distributes positive gains across a substantially broader set of prosumer-time observations. By contrast, the zero median obtained for the nucleolus indicates that at least half of the observations receive no strictly positive gain beyond their stand-alone payoff. 

\begin{figure}[h]
    \centering
    \includegraphics[width=1\linewidth]{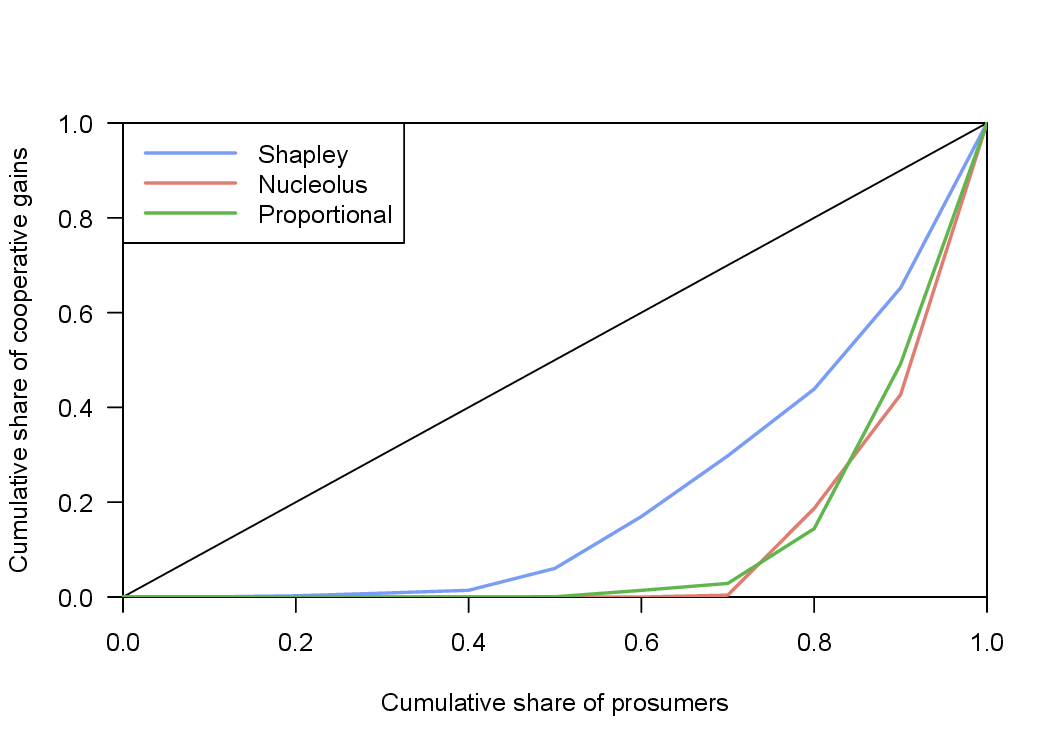}
    \caption{Lorenz curves of the cooperative gains $g_{i,t}^r=x_{i,t}^r-v_t(i)$  under the settlement rules at $t=20$.
    }
    \label{fig:Lorenz_t20}
\end{figure}

 To further illustrate the differences between the allocation methods, Figure \ref{fig:Lorenz_t20} compares the distribution of the cooperative gains generated by the three rules. The Lorenz curve plots the cumulative proportion of total cooperative gains received by the cumulative proportion of prosumers, ordered by increasing gain. The diagonal represents an equal division of the aggregate cooperative surplus among all prosumers. In the considered instance, the Lorenz curve associated with the Shapley value lies closer to the equality line. The corresponding distribution of cooperative gains is therefore less concentrated than those generated by the nucleolus and the proportional rule. The Gini coefficient measures the area between the Lorenz curve and the diagonal line of equal distribution. A curve closer to the diagonal produces a lower Gini coefficient and represents a more even distribution of cooperative gains. At the time instant shown in Figure~\ref{fig:Lorenz_t20}, the Gini coefficients are 0.57 for the Shapley value, 0.77 for the nucleolus, and 0.76 for the proportional rule.

 %In this context, the Lorenz curves are constructed from the cooperative gains $x_i-v_t(i)$, rather than from the total allocations, and the diagonal line represents the assignment $v_t(i)+\frac{v_t(\mathcal N)-\sum_{i\in{\mathcal N}}v_t(i)}{n}$; that is, everyone receives their individual worth, and the remaining amount is distributed equally among everyone. The Shapley value distributes this amount more equitably, whereas the other two allocations result in greater inequality among prosumers.

To further assess how the surplus is distributed among prosumers, we compute the Gini coefficient of the gains generated by each allocation rule at every time instant  with a strictly positive cooperative surplus. Averaging the time-specific coefficients over the analyzed day yields values of 0.63 for the Shapley value, 0.81 for the nucleolus, and 0.72 for the proportional rule. Moreover, the Shapley value produces the lowest Gini coefficient at every time instant, indicating that, throughout the analyzed day, it consistently distributes the surplus more evenly among prosumers than the other two mechanisms. By contrast, the relative performance of the nucleolus and the proportional rule varies over time: the nucleolus yields the lower Gini coefficient in some periods, whereas the proportional rule does so in others. These results reinforce the previous finding that the Shapley value distributes positive gains across a broader set of prosumer-time observations.

\begin{figure}[h]
    \centering
    \includegraphics[width=1\linewidth]{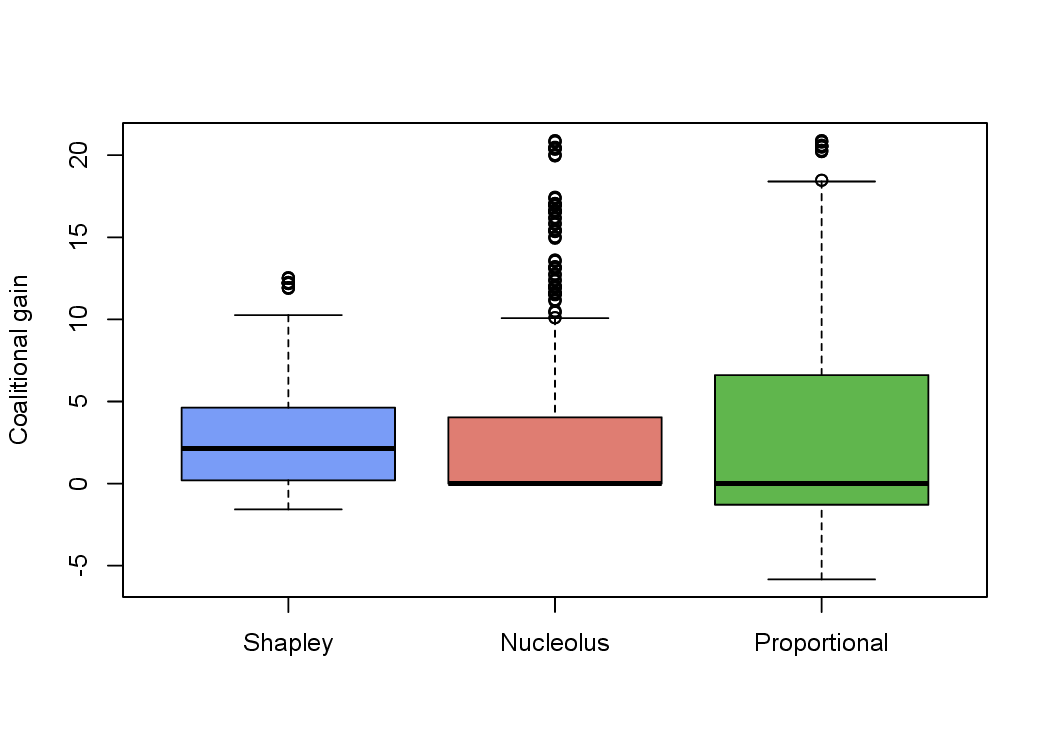}
    \caption{Boxplot of the distribution of coalitional gains $g_{S,t}^r=x_{S,t}^r-v_t(S)$ over all coalitions $S\in 2^{\mathcal N}$ under the settlement rules at $t=20$.}
    \label{fig:Boxplot_t20}
\end{figure}

Now, we study the effects of the property of coalitional rationality. We compute, for every coalition $S\in 2^{\mathcal N}$, the coalitional gain $g_{S,t}^r=x_{S,t}^r-v_t(S)$. A negative value indicates that coalition $S$ could obtain a greater aggregate payoff by settling separately and therefore identifies a violation of coalitional rationality. Figure~\ref{fig:Boxplot_t20} presents a boxplot of the distribution of these gains for the three allocation rules at $t=20$. The nucleolus is the only rule that satisfies coalitional rationality for every coalition in this instance. Thus, no coalition receives less than its stand-alone worth. Some coalitions receive exactly their stand-alone worth, as reflected by the zero median coalitional gain. %A median coalitional gain equal to zero indicates that at least half of the coalitions obtain a gain no greater than zero, up to numerical tolerance. For the nucleolus, which satisfies coalitional rationality, these observations correspond to coalitions lying on the boundary of the core. 
For the proportional rule, some coalitions exhibit negative gains and therefore have incentives to deviate. On the other hand, the median Shapley value is greater than zero, so most coalitions improve their allocation.

The day-wide analysis also reveals an important distinction between benefiting most coalitions and satisfying coalitional rationality. In the time instants with heterogeneous deviations, approximately 80\% of the coalitions receive, under the Shapley value, an allocation greater than or equal to $v_t(S)$. Moreover, its median coalitional gain is positive and equal to 0.44, which means that at least half of the coalitions obtain a strictly positive benefit from participating in the grand coalition. However, the Shapley allocation lies outside the core at every heterogeneous-deviation time instant, since at least one coalition always receives less than $v_t(S)$. The proportional rule also fails to satisfy coalitional rationality during the analyzed day: 68\% of the coalition-time observations receive an allocation greater than or equal to their coalition worth. Its median coalitional gain is zero, indicating that the center of its distribution lies on the boundary between coalitional gains and losses. By contrast, the nucleolus produces no negative coalitional gains and therefore satisfies coalitional rationality at every time instant. Its median gain is also zero; however, since all nucleolus gains are nonnegative, this result indicates that at least half of the coalitions receive exactly their stand-alone worth rather than a strict participation gain.

An additional distinction between the three methods lies in the additivity of the Shapley value. Since it is additive, if the final goal is to compute the total allocation throughout the day, we can apply it directly to the daily LEM game. In this case, the resulting allocation coincides with the sum of the allocations computed at each time instant of the RTLEM games, which avoids the need to compute both the daily Shapley value and the collection of time-specific Shapley values separately. In contrast, the nucleolus is not additive in general, and the daily allocation of the proportional rule is obtained by summing the settlements computed separately for each time instant. Table~\ref{tab:additive} confirms the additivity of the Shapley value compared with the nucleolus.

\begin{table}[h] \centering \footnotesize \sisetup{ table-align-text-post = false,   round-mode = places, round-precision = 2 } \begin{tabular}{c S[table-format=-2.2]  S[table-format=-2.2] S[table-format=-2.2]} \toprule Prosumer & {$\sum_{t\in\mathcal T}Sh_i(v_{t})=Sh_i(v)$} & {$\sum_{t\in\mathcal T}Nu_i(v_{t})$} & {$ Nu_i(v)$} \\ 
\midrule 
1 & 235.32 & 236.51 & 240.52\\
2 & 39.08 & 44.11 & 49.15 \\ 
3 & 374.35 & 374.23 & 370.79\\
4 & 254.77& 248.75 & 242.95\\
5 & -80.33& -80.33 & -85.23\\ 
6 & 262.85 & 261.08 & 260.59\\ 
7 & -131.56 & -143.19& -153.51 \\
8 & 96.51 & 91.68 & 96.81 \\ 
9 & 163.71 & 176.46& 182.30 \\
10 & -1120.17 & -1114.79 & -1109.85 \\ \bottomrule \end{tabular} \caption{Comparison between time-aggregated and directly computed daily Shapley and nucleolus allocations.} \label{tab:additive} \end{table}

\subsection{Discussion and Practical Implications}

From an implementation perspective, the choice of a settlement rule involves a trade-off between interpretability, distributional objectives, coalitional stability, and computational effort. The proportional rule can be explained through individual deviations and proportional responsibility, making it comparatively transparent. However, although it is efficient and individually rational, it may fail to satisfy coalitional rationality.

The Shapley value implements a contribution-based allocation principle by rewarding prosumers according to their expected marginal contributions across all possible coalition-formation orders. This interpretation may be appealing when the community wishes to recognize each participant’s contribution to the cooperative surplus. Besides, the Shapley value produces the most even distribution of cooperative gains: it assigns strictly positive gains to a broader proportion of prosumer-time observations and yields the lowest Gini coefficient at every time instant. Moreover, its additivity across games provides an important implementation advantage: the daily Shapley allocation can be computed directly from the aggregated daily game, without having to compute a separate Shapley allocation at every time instant. Nevertheless, the Shapley value is not guaranteed to belong to the core.

The nucleolus follows a different principle. It selects a coalitionally stable allocation by lexicographically minimizing coalition excesses. Since every RTLEM game is balanced, the nucleolus belongs to the core and therefore guarantees coalitional rationality. Among the three mechanisms considered, it is the only rule with this guarantee for every RTLEM instance.

Accordingly, no rule dominates the others under all criteria. The proportional rule prioritizes simplicity and interpretability. The Shapley value provides temporal-aggregation consistency, and produces the broadest and least concentrated distribution of cooperative gains in the analyzed data. The nucleolus prioritizes coalitional stability and is the only rule with a general core-membership guarantee.

\section{Conclusions}
\label{sec:concl}
This paper introduced a cooperative game-theoretical framework for financial settlements in CB-LEMs. By modeling the settlement process as a cooperative game, we compared different settlement rules, including the proportional rule, the Shapley value and the nucleolus of the RTLEM game.
The analysis leads to several key conclusions. First, the RTLEM game is balanced, which guarantees the existence of coalitionally stable allocations and ensures that the nucleolus belongs to the core. The Shapley value remains efficient and individually rational, but, because the RTLEM game is not necessarily convex, the Shapley value is not guaranteed to satisfy coalitional rationality. Second, while the proportional rule is computationally simple, it may fail to align coalitional incentives with community goals, whereas the nucleolus provides a more robust foundation for sustainable participation. Finally, from a computational point of view, the exact computation of the Shapley value and nucleolus may become challenging as community size increases, although additivity properties can simplify temporal aggregation. The case study based on a residential energy community in Amsterdam illustrated the practical implications of the proposed framework, showing how different allocation rules impact the distribution of costs and benefits under real-world conditions. %To support this calculation, we utilized TUGLab \cite{TUGlab}, an open source R package developed for the analysis and implementation of cooperative game-theoretic tools. %In particular, the methodology for computing the Shapley value is detailed in \cite{Computation_shapley}, where, based on the linearity of the solution, an algorithm to calculate this value is presented. %The algorithm is also published in \cite{randomarrival}.
In future work, we aim to explore the application of sampling algorithms and parallel computation to extend our framework to larger communities. %In such cases, to compute the Shapley value, sampling-based approximation methods, such as Monte Carlo sampling of marginal contributions, will be necessary to efficiently estimate Shapley allocations.% Some examples are \cite{mann1960values,owen1972multilinear,maleki2015addressing,castro2017improving}. There are also studies that approximate the Shapley value \cite{han2021estimation} and the  nucleolus \cite{han2020scaling,han2019improving} in the context of energy communities.

\vspace{1em}

\bibliographystyle{IEEEtran}

\bibliography{ref/references}
\end{document}